\documentclass[a4paper,useAMS,usenatbib]{mnras}
\usepackage{graphicx}
\usepackage{color,ulem}

\usepackage[usenames,dvipsnames,svgnames,table]{xcolor}

\def\pc{\mbox{\,pc}}

\def\Myr{\mbox{\,Myr}}

\def\kms{\mbox{\,km s}^{-1}}

\def\mucrit{\mu_{{\rm crit}}}
\def\taudif{\tau_{{\rm dif}}}

\def\pcc{\mbox{\,cm}^{-3}}

\def\deg{{^\circ}}

\newcommand{\Linf}{L_{\rm inf}}
\newcommand{\Rinf}{R_{\rm inf}}
\newcommand{\K}{{\rm K}}

\title[Magnetic field structure in molecular filaments.]
      {The magnetic field structure in molecular cloud filaments.}

\author[G\'omez, V\'azquez-Semadeni \& Zamora-Avil\'es]
{
Gilberto C. G\'omez$^{1}$ \thanks{E-mail: g.gomez@irya.unam.mx},
Enrique V\'azquez-Semadeni$^{1}$, and Manuel Zamora-Avil\'es$^{1,2}$
\\
$^{1}$Instituto de Radioastronom\'ia y Astrof\'isica,
      Universidad Nacional Aut\'onoma de M\'exico, Apdo. postal 3-72,
      Morelia Mich. 58089, M\'exico \\
$^{2}$Department of Astronomy, University of Michigan, 500 Church Street,
      Ann Arbor, MI 48105, USA}

\begin{document}

\date{Accepted . Received ; in original form }

\pagerange{\pageref{firstpage}--\pageref{lastpage}} \pubyear{}

\maketitle

\label{firstpage}

\begin{abstract}

  We explore the structure of magnetic field lines in and around
  filaments in simulations of molecular clouds undergoing global,
  multi-scale gravitational collapse. In these simulations, filaments
  are not in a static equilibrium, but are long-lived flow structures
  that accrete gas from their environment and direct it toward clumps
  embedded in the filament or at the nodes at the conjunction with
  other filaments. In this context, the magnetic field is dragged by
  the collapsing gas, so its structure must reflect the flow that
  generates the filament. Around the filament, the gas is accreted
  onto it, and the magnetic lines must then be perpendicular to the
  filament. As the gas density increases, the gas flow changes
  direction, becoming almost parallel to the filament, and magnetic
  lines also tend to align with it. At the spine of the filament,
  however, magnetic lines become perpendicular again since they must
  connect to lines on the opposite side of the filament, resulting
  in``U''-shaped magnetic structures, which tend to be stretched by
  the longitudinal flow along the filament. Magnetic diffusive
  processes, however, allow the gas to continue to flow. Assuming a
  stationary state in which the ram pressure of the flow balances the
  magnetic tension, the curvature of the field lines is determined by
  the diffusion rate. We derive an expression relating the curvature
  of the field lines to the diffusive coefficient, which may be used
  to observationally determine the nature of the diffusive process.

\end{abstract}


%%%%%%%%%%%%%%%%%%%%%%%%%%%%%%%%%%%%%%%%%%%%%%%%%%%%%%%%%%%%%%%%%%%%%%%%%%%%%%%%
\section{Introduction}\label{sec:Intro}

The ubiquity of filaments in molecular clouds has been clearly
established by numerous infrared continuum and molecular line
observations \citep{bally1987, myers2009, and2010, mol2010, arz2011,
palm2013, kirk2013, peret2014, li2016, rivera2016}.
However, the internal structure and kinematics of
the filaments is still debated. 
A class of published models assumes that the
filaments are in hydrostatic equilibrium \citep[e.g.,] []
{sto1963, ost1964, inut1992, fis2012, burge2016}, but this leaves open
the question of how the filaments were formed in the first
place and how they reached such equilibrium.  Another proposed model
assumes that the filaments form as a consequence of turbulent
compressions in the clouds \citep[e.g.,] [] {padoan2001, auddy2016}.
Finally, yet another class of models assumes that the filaments form as
part of the global gravitational collapse of
a molecular cloud as a
whole, and therefore the filaments radially accrete material from their
parent clouds \citep{heitsch2013a, heitsch2013b, hennebelle13}. However,
neither of these models accounts for the observation that the gas
appears to be flowing along the filaments towards the cores and ``hubs''
located within the filaments \citep[e.g.,] [] {schnei2010, kirk2013,
peret2014}.

A different scenario was proposed by \citet[] [hereafter GV14;
see also \citealt{smith2016}]
{gom14}. Based on numerical simulations of molecular cloud formation and
subsequent collapse in which filaments developed self-consistently,
these authors suggested that the filaments are
flow structures, accreting gas from the cloud and redirecting
it toward the star-forming clumps that are formed both within the
filaments and at their intersections with other filaments.
%\citet{gom14} 
In this picture, the filaments are out-of-equilibrium,
river-like structures, along which the gas from
a cloud flows down the gravitational potential
towards a clump located at the potential minimum.
The filament is, simply, the locus of
this flow along a local trough in the potential. In this
scenario, the filaments arise from the strongly anisotropic global
collapse of the cloud, which naturally forms sheets and filaments
\citep{lin1965}, growing from small perturbations, and then reaching
approximately stationary regimes for as long as the gas supply from the
cloud remains (GV14). Thus, their growth and observed dynamical nature
are naturally explained. It is important to remark that the flow
observed by GV14 along the filaments did not develop internal shocks nor
strong turbulence, the flow being relatively laminar.

Magnetic fields have also been observed associated to molecular clouds
and embedded filaments, often appearing roughly perpendicular
to the filaments \citep[e.g.,] [] {goo1992,cha2011, palm2013, planckXXXV}.
Their impact in the molecular cloud formation and evolution is
still a subject of debate, although a frequent interpretation
is that the magnetic field is somehow responsible for the formation of
the filament, by allowing the gas to flow towards it. However, this
interpretation does not provide an answer to the question of what is
driving the gas to flow along the field towards the filament.
In this regard, \citet{VS+11} presented simulations of cloud
formation in the colliding flow scenario in the presence of
self-gravity and magnetic fields oriented parallel to the colliding
flows. These simulations showed
that the evolution is similar to that of the non-magnetic
case, because the cloud increases its mass-to-flux ratio (M2FR) by
accretion along the field lines, eventually becoming magnetically
supercritical and proceeding to global collapse.
The cloud in those
simulations consists of high-density, high-M2FR clumps surrounded by
low-density, low-M2FR bubbles.
These low-density bubbles are buoyant and allow the dense clumps to
percolate to the center of the cloud
``through a process that appears as the macroscopic-scale analogue of AD
(ambipolar diffusion) (\dots)
so that the clouds evolve
towards a segregated state with low M2FR in their periphery and high
M2FR towards their centre''.

Since the evolution of the cloud presented in \citet{VS+11} is similar
to that observed in nonmagnetic simulations of molecular cloud
formation, including the formation of filaments
\citep[e.g.,][]{VS+07, HH08}, it is necessary to determine whether the
filaments formed in magnetized simulations also behave as the flow
structures mentioned earlier and, if so, how
is the magnetic field affected by the accretion
towards and along a filament formed in a cloud undergoing global
gravitational collapse. In this contribution we explore this question
in one of the MHD simulations presented by \citet{ZA+17}.  In
\S\ref{sec:model} we briefly describe the simulation.  In
\S\ref{sec:mag_structure} we discuss the resulting magnetic field
structure associated to the flow around a filament in the simulation.
%In \S\ref{sec:evolution} we discuss the temporal evolution
%of the filament.
In \S\ref{sec:diffusion} we discuss how
the magnetic field geometry may be used to explore the
dominant diffusive process in the medium.
Finally, in \S\ref{sec:discusion} we summarize our conclusions.

%%%%%%%%%%%%%%%%%%%%%%%%%%%%%%%%%%%%%%%%%%%%%%%%%%%%%%%%%%%%%%%%%%%%%%%%%%%%%%%
\section{The Numerical Model}\label{sec:model}

The filament studied here comes from the MHD simulation presented in 
\cite{ZA+17} labeled B3J, which is aimed at studying the effect of 
magnetic fields on the production of turbulence through various 
instabilities during the formation of MCs by converging flows and the 
subsequent star formation activity. It was performed using the Eulerian 
adaptive mesh refinement FLASH (v2.5) code \citep{Fryxell+00} to
obtain three-dimensional, self gravitating, ideal--MHD simulations,
including heating and cooling processes, and a Jeans
refinement criterion. For more details, we
refer the reader to \citet{ZA+17}.

%\jade{¿Y si quitamos este párrafo?} {\magenta [De acuerdo con quitar
%\'este y el que sigue, a\~nadiendo la frase que puse en el p\'arrafo
%anterior.]}  For the dynamical mesh refinement, the so--called Jeans
%criterion is used, where the local Jeans length is resolved with at
%least ten grid cells in order to prevent spurious fragmentation
%\citep{Truelove+97}. When the maximum refinement level is reached in a
%given cell, a sink particle can be formed when the density in this
%cell exceeds a threshold number density, $n_{\rm thr}$, among other
%standard sink-formation tests
%\citep{Federrath+10}.

%\jade{¿Y este también?}
%The heating and cooling rates are calculated using the analytic fits by 
%\citet{KI02} for the heating and cooling functions \citep[see 
%also][]{VS+07}, which are based on  the thermal and chemical 
%calculations considered by \citet{Wolfire+95} and \citet{KI00}. For a 
%more detailed description of the numerical model we refer the reader to 
%\cite{ZA+17}.

In this simulation, the numerical box, of dimensions $L_x=256 \pc$ 
and $L_y = L_z=128  \pc$, is initially filled with warm neutral gas at 
a uniform density $n = 2  \pcc$ (with a mean molecular weight $\mu = 
1.27$) and constant temperature of $1450  \K$, which correspond to 
thermal equilibrium.
The initial velocity field contains turbulent 
fluctuations with a Burger's type spectrum
corresponding to a rms Mach number of $0.7$ (with respect to the 
isothermal sound speed of $3.0 \kms$). On top of this turbulent field, 
we add two cylindrical streams, each of radius $\Rinf=32 \pc$ and 
length $\Linf = 112 \pc$ along the $x$--direction, moving in opposite 
directions at a moderately supersonic velocity of $7.5 \kms$.

The numerical box is permeated with an initially uniform magnetic field 
of $3 \mu$G parallel to the inflows.\footnote{Note that the 
corresponding mass-\-to-\-flux ratio is 1.59 times greater than the 
critical value, $\mucrit = 0.16/\sqrt{G}$ \citep{NN78}, so that our 
cloud is mag\-ne\-ti\-cally supercritical.} Unlike the simulations by 
\cite{ZA+17} (which have a maximum resolution of $\Delta x = 3.1 \times 
10^{-2} \pc$), here we allow the simulation to refine eight levels 
more to reach a maximum resolution of $1.2 \times 10^{-4} \pc$. The 
corresponding sink formation density threshold is $n_{\rm thr}=2.1 
\times 10^{9} \pcc$.

Although it has been stated that a magnetic field perpendicular to
the inflows, i.e. parallel to the shock surface, may inhibit the
formation of a cloud \citep{inoue09}, we consider such a configuration as
unlikely to happen in reality.
The colliding flow scenario used here is thought to be
representative of the gas flow in and around spiral arms, where
magnetic field is parallel to the large-scale gas flow \citep[see
for example, fig. 8 in][]{gom2004}.

\begin{figure}
  \begin{center}
    \includegraphics[width=0.45\textwidth]{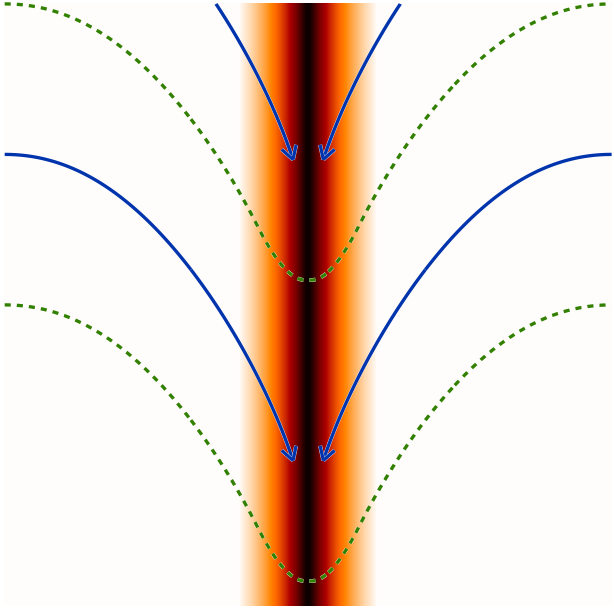}
  \end{center}
  \caption{Simplified picture of the ``U''-shaped magnetic line geometry
    due to the accretion toward and along the filament.
    Colours indicate the projected column density of the filament,
    while the solid lines show the flow of gas and the dashed lines show
    the magnetic field lines, dragged by the accretion flow.
    The filament column density profile and the velocity field are an
    approximation to the fits presented in GV14.
  }
  \label{fig:cartoon}
\end{figure}

%%%%%%%%%%%%%%%%%%%%%%%%%%%%%%%%%%%%%%%%%%%%%%%%%%%%%%%%%%%%%%%%%%%%%%%%%%%%%%%
\section{Magnetic Field Structure}
\label{sec:mag_structure}

As described in GV14, as the cloud collapses as a whole,
filaments are formed in the regions where the gas accretion turns
from two-dimensional (sheet-like accretion to the filaments from
their environment) to one-dimensional (along the filaments to
cores).
This accretion occurs where the magnetic field is weaker
and, therefore, it is dragged by the gas flow.
A simple picture of the resulting magnetic field line geometry
is presented in Fig. \ref{fig:cartoon}.
Around the filament, the magnetic field must be mainly perpendicular
to the filament since the accretion flow has that direction.
As the gas velocity turns and the flow becomes dominated by the
longitudinal motion along the filament,
the magnetic lines must also develop a component parallel to the filament.
At some point, the magnetic lines must turn back and connect
with the (perpendicular) lines on the other side of the filament.
So, at the filament spine, the magnetic field must be perpendicular
to the filament again, resulting in a ``U''-shaped magnetic line across the
filament: perpendicular in the surroundings, then oblique, and
perpendicular again in the center.
Of course, in this high-density region the increasing magnetic tension
and the opposite magnetic polarities must play a role, either
through magnetic reconnection or diffusion
(\S \ref{sec:diffusion}. See also \citealt{hennebelle13}).

\begin{figure}
  \includegraphics[width=0.5\textwidth]{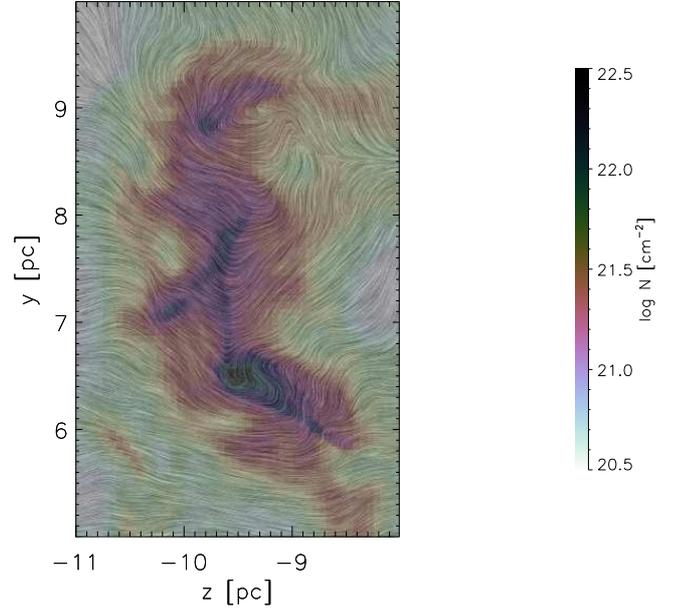}
  \caption{Column density of a filament with its magnetic field
    direction, $13.1\Myr$ after the beginning of the simulation.
    Coordinates are with respect to the centre of the simulation
    box.
    The overlaid ``drapery'' pattern represents the (density-weighted) mean
    magnetic field in the $z-y$ plane,
    visualized using the line-integral-convolution technique \citep{cab93}.
    The ``U''-shaped magnetic structures due to the gas accretion are
    apparent.
  }
  \label{fig:fil1}
\end{figure}
\begin{figure}
  \includegraphics[width=0.5\textwidth]{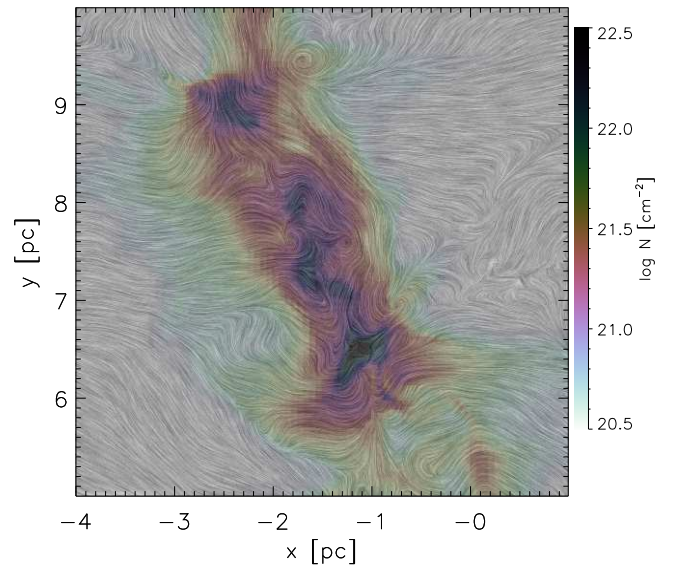}
  \caption{Same region as in Fig. \ref{fig:fil1},
    but with column density and mean magnetic field integrated along
    the $z$ direction.
    In contrast with that figure, the ``U''-shaped magnetic
    structures are not visible.
  }
  \label{fig:fil1-xy}
\end{figure}

Fig. \ref{fig:fil1} shows a filament of the simulation.
Colours show the column density of gas, while the ``drapery'' pattern
indicates the direction of the (density-weighted) mean magnetic field.
The above described ``U''-shaped magnetic lines associated with the
filament are clearly visible in the vertical segment at
$(z,y) \approx (-9.5,7)\pc$ and the slanted segment at
$(z,y) \approx (-9,6)\pc$.
They are also visible, although not as clearly, in the segment near
the top of the filament, at $(z,y) \approx (-9.5,8)\pc$.
Please notice that near the bottom of the filament,
at $(z,y) \approx (-9,6)\pc$,
the ``U''-shaped lines appear to point away from the embedded clump.
This do not contradict the above described picture since
the molecular cloud as a whole is experiencing gravitational
collapse. Therefore, the structures contained within the cloud
also drag the magnetic field and generate magnetic
``U''-structures associated to this large scale flow.

Obviously, the scenario described in Fig. \ref{fig:cartoon} is an
idealized one and an axisymmetric version would be difficult to achieve.
As an example, Fig. \ref{fig:fil1-xy} shows another projection
of the same filament.
In this case, the ``U''-shaped structures are not as clear as in the
Fig. \ref{fig:fil1}.
A possible reason for this is the following:
The near-pressureless collapse of
a molecular cloud means that density structures collapse
along the shortest dimension first, going from a (three-dimensional)
ellipsoid to a (two-dimensional) sheet and further to a
(one-dimensional) filament.
So, the accretion toward the filament will not be axisymmetric, but
from a planar environment.
In this scenario, it is not surprising that the magnetic field lines,
dragged by the flow, will adopt a geometry that reflects this
dimensionality of the accretion flow.
So, the ``U''-shaped magnetic lines associated with the filaments
might be observable along some directions only,
but their absence may not be considered as
evidence of lack of flow along the filament.

%%%%%%%%%%%%%%%%%%%%%%%%%%%%%%%%%%%%%%%%%%%%%%%%%%%%%%%%%%%%%%%%%%%%%%%%%%%%%%%
\subsection{Filament evolution}\label{sec:evolution}

\begin{figure*}
  \includegraphics[width=\textwidth]{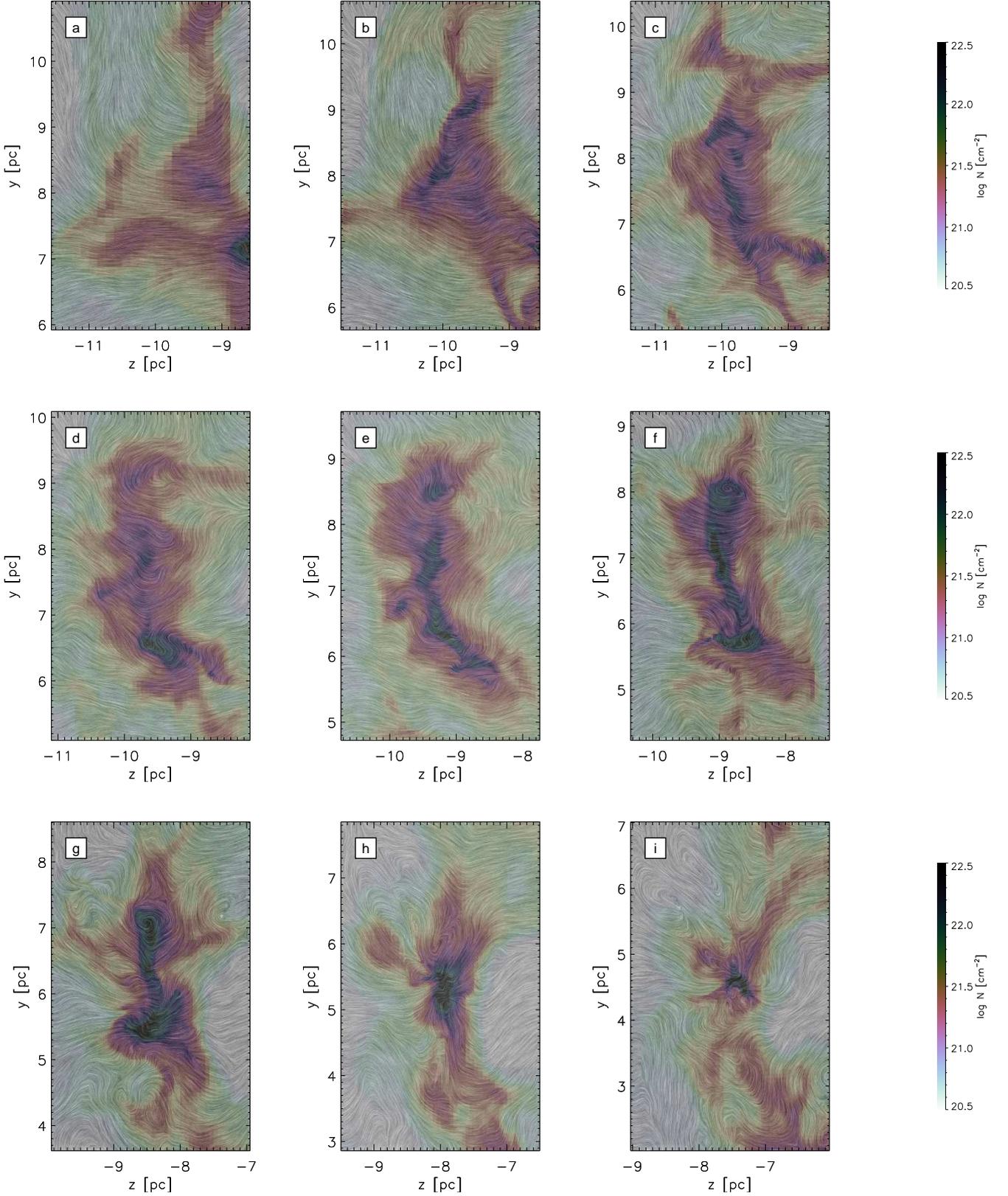}
  \caption{Evolution of the filament and its associated magnetic structure.
  Times shown correspond to $10.8$ (panel {\it a}) through $16.4\Myr$
  (panel {\it i}) from the start of the simulation, in $0.7\Myr$ increments.
  Although all panels show regions of the same size, their positions
  in the simulation box change in order to follow the evolution of
  the filament as it moves following the large-scale gravitational
  collapse.
  }
  \label{fig:evol}
\end{figure*}

Figure \ref{fig:evol} shows the time evolution of the filament along
$5.6\Myr$.
The now familiar ``U''-shaped magnetic structures are visible
before the filament is apparent in density (panel {\it a}) since,
on the one hand,
the compressive flow that originates the filament must predate it,
and on the other, the cloud as a whole is already undergoing global
collapse, while at the same time
some collapse centers are already present, as
filaments and cores are observed to form simultaneously in the
simulations \citep{camacho2016}.
Therefore, the accretion towards the (future) filament
and towards the embedded core is already happening and so it should
drag the magnetic field with it.

These broad initial ``U''-structures become narrower as the filament's
density increases and the accretion flow becomes better defined.
Eventually, smaller collapse centers appear within the filament
(as described in GV14),
leading to a more complex velocity and magnetic structure.
$14.3\Myr$ into the simulation (panel {\it f}) the filament-core
self-gravity start to dominate the small scale evolution and,
while the filament as a whole keeps falling into the cloud's center,
it rapidly collapses longitudinally.
This is reflected in less prominent large-scale ``U''-structures
(although some are still visible near the larger cores).
At $16.4\Myr$ (panel {\it i}),
the filament as a whole collapses into a large core,
with some low-column density structures remaining in the
surroundings.
By this time, stellar feedback (not included in this simulation)
should have an impact on the low density gas around the core
\citep{colin13}, and so this aspect of the simulation becomes less
reliable.

%%%%%%%%%%%%%%%%%%%%%%%%%%%%%%%%%%%%%%%%%%%%%%%%%%%%%%%%%%%%%%%%%%%%%%%%%%%%%%%
\section{Magnetic Diffusion}\label{sec:diffusion}

As the gas flow along the center of the filament drags the magnetic field,
the increasing curvature of the ``U''-shaped lines means that the
magnetic forces can no longer be dismissed.
If we assume that diffusion effects separate the gas flow from the
magnetic field lines, we may connect observables with the
characteristics of the flow along the filament or with the parameters
of the diffusion process at work.

%\begin{figure}
%  \begin{center}
%    \includegraphics[width=0.45\textwidth]{figures/diffusion.eps}
%  \end{center}
%  \caption{Definition of variables used in \S\ref{sec:diffusion}
%    (see text).
%    {\bf ?`Necesitamos esta figura? Nos empieza a faltar espacio}
%  }
%  \label{fig:diff}
%\end{figure}

Consider the simple picture of the filament and associated magnetic
lines shown in Fig. \ref{fig:cartoon}.
At the filament's spine,
the shape of the magnetic line will be defined by a balance between
magnetic tension and the ram-pressure due to the flow along the
filament:

\begin{equation}
  \frac{\rho v_A^2}{R_c} = \frac{\rho v_l^2}{2L},
  \label{eq:force_balance}
\end{equation}

\noindent
where $\rho$ is the gas density, $v_A$ is the Alfv\'en velocity,
$R_c$ is the line curvature,
$v_l$ is the gas velocity along the filament, and
$L$ is the filament half-width.
%(see fig. \ref{fig:diff}).
If $R_c > L$, we may rewrite equation (\ref{eq:force_balance}) in
terms of easier to measure quantities.
Let $l$ be the depth of the ``U''-shaped line along the filament.
Then, the angle between the magnetic line and the direction perpendicular
to the filament, $\alpha$, is related to the above defined lengths
by $\tan (\alpha) = l/L$.
In turn, $l$ is related to the curvature radius by $R_c^2 = (R_c-l)^2 + L^2$.
Simple algebraic manipulation leads to $R_c = L/\sin(2\alpha)$, and
thus equation (\ref{eq:force_balance}) becomes,

\begin{equation}
  \left( \frac{v_l}{v_A} \right)^2 = 2 \sin( 2\alpha).
\end{equation}

The physics of the diffusive process is reflected
in the length the gas moves before the magnetic field decouples from
the flow, so $l = v_l \taudif$, where $\taudif$ is
the diffusion timescale.
Then, we may use the observables $\alpha$, $L$, $v_l$, and $v_A$
to estimate $\taudif$ and so explore the magnetic diffusion
process at hand.
For example, \citet{Lazarian99} explored the relation between
turbulence parameters and fast reconnection in the ISM.
By measuring the shape of the magnetic lines and the flow associated
to a molecular filament,
we may be able to estimate the turbulent characteristics
of the flow within.

Conversely, we may assume a diffusion process and then estimate some
observable quantity.
For example, if we assume that the source of the magnetic diffusion
is ion-neutral friction, then $\taudif$ is known
\citep[e.g.][]{hennebelle13}, and we may use $\alpha$, $L$, and $v_A$
to obtain an estimation of the accretion along the filament $v_l$.
The effects of observational resolution and inclination with respect
to the plane of the sky will be explored in the future.

%%%%%%%%%%%%%%%%%%%%%%%%%%%%%%%%%%%%%%%%%%%%%%%%%%%%%%%%%%%%%%%%%%%%%%%%%%%%%%%
\section{Discussion}\label{sec:discusion}

Observations of magnetic fields in molecular cloud
filaments show that the field is
perpendicular to the filaments in the surrounding and
central regions \citep{goo1992,palm2013}.
This geometry is usually interpreted as the field being
perpendicular to the filament, which is then taken as evidence for
the flow being dominated by the field \citep{sugitani2011,li2013}.
The picture described in this article, in which the flow is dominated by gas
accretion and the field is dragged with the flow,
is not inconsistent with the observations, since the
magnetic field is harder to observe in the intermediate density
regions where the lines turn and become aligned by the flow with the
filament.
Observational resolution, or lack thereof, would lead to less
pronounced observed ``U''-shaped lines, in addition to a lower
degree of polarization in the emission.
We will explore this observational effects in a future work.

Magnetic structures qualitatively similar to the ones
described here are visible in
the Planck maps of magnetic fields
associated to molecular clouds.
For example, fig. 1 in \citet{planckXXXV} shows the column density
and magnetic field direction towards the Taurus molecular cloud.
At galactic coordinates $(l,b) \sim (168\deg,16.2\deg)$,
the magnetic field lines follow
the ``U''-shape we associate with the accretion flow around and
along a filament in a globally collapsing cloud.
Further left in the same figure,
the magnetic lines are still bent for some $4\deg$ along galactic
longitude.
Although the reader should keep in mind that the physical scales
in the Taurus filaments and our simulation are quite different,
the similarity of that figure and our Fig. 5({\it e}) is striking
and it is easy to speculate if a similar mechanism is at work.
Curved magnetic fields have also been observed in the
main filament of Serpens South, which have been
associated to gravitational contraction along the filament
\citep{sugitani2011}.

An important assumption for this work is that the gas dynamics of
the collapsing cloud is dominated by self-gravity, and so the
evolution might be considered pressureless.
\citet{VS+11} showed that the magnetically supercritical regions
percolate away from the cloud, thus allowing for the collapse of the
subcritical regions, so the presence of a magnetic field should not
alter significantly the large scale evolution of the cloud.
In the picture presented here, the gas accretion flow along the
filament will be affected by the magnetic tension in the dense
central regions of the filament, since the
``U''-shaped lines counteract the gravitational flow toward the
collapsing regions embedded within the filaments until the magnetic
diffusion allows the flow to continue.
Thus, the accretion along the filament which feeds the star forming
cores will be connected to the details of the diffusive process
happening in the central regions of the filaments.
Since ambipolar diffusion is too slow to play an
important part in the filament dynamics, turbulent reconnection
appears to be an interesting possibility to separate the gas flow
from magnetic stresses.
Since this reconnection is related to the turbulent
characteristics of the fluid, the study of ``U''-shaped magnetic
field lines might be useful to study the nature of unresolved
motions in the central parts of molecular filaments.

%%%%%%%%%%%%%%%%%%%%%%%%%%%%%%%%%%%%%%%%%%%%%%%%%%%%%%%%%%%%%%%%%%%%%%%%%%%%%%%
\section*{Acknowledgments}

We thank B. Bukhart, I. Ristorcelli and J. Soler for useful comments
during the development of this project.
This work received financial support from UNAM-DGAPA PAPIIT grant IN100916
to G.C.G. and CONACyT grant 255295 to E.V-S.
M.Z.-A. acknowledges CONACyT for a postdoctoral fellowship at
University of Michigan.

%%%%%%%%%%%%%%%%%%%%%%%%%%%%%%%%%%%%%%%%%%%%%%%%%%%%%%%%%%%%%%%%%%%%%%%%%%%%%%%

\label{lastpage}

\end{document}